\documentclass[aps,prb,twocolumn,superscriptaddress,showpacs]{revtex4}

\usepackage{graphicx}
\usepackage{bm}

\renewcommand{\vec}[1]{\bm{#1}}

\begin{document}

\title{Dynamics of anisotropic spin dimer system in strong magnetic field}

\author{A. K. Kolezhuk}
\affiliation{Institute of Magnetism, National Academy of Sciences and
Ministry of Science and Education, 
03142 Kiev, Ukraine }
\affiliation{Institut f\"ur Theoretische Physik, Universit\"at
Hannover, 
30167 Hannover, Germany}

\author{V. N. Glazkov} 
\affiliation{P. L. Kapitza Institute for Physical Problems RAS, 117334
 Moscow, Russia}
\affiliation{Commissariat \`a l'Energie Atomique, DSM/DRFMC/SPSMS, 
38054 Grenoble, Cedex 9, France}

\author{H. Tanaka} 
\affiliation{Research Center for Low Temperature Physics, Tokyo
Institute of Technology, Meguro-ku, Tokyo 152-8551, Japan }

\author{A. Oosawa}
\affiliation{Advanced Science Research Center, 
Japan Atomic Energy Research Institute, Tokai, Ibaraki
319-1195, Japan}

\date{\today}

\begin{abstract}

Recently measured high-field ESR spectra of the spin dimer material
 $\rm TlCuCl_{3}$ are described within the
framework of an effective field theory. A  good agreement between the theory and
experiment is achieved, for all geometries and in a wide field range, under the
assumption of a weak  anisotropy of the interdimer as well as
intradimer exchange interaction.

\end{abstract}

\maketitle

\section{Introduction}

Gapped spin systems in high magnetic field have attracted much attention recently,
both from the theoretical and experimental side. In absence of the field the system is
supposed to have a singlet ground state and a finite gap $\Delta$ to the lowest
excitation, typically in the triplet sector.  When the field is increased beyond the
critical value $H_{c}$ necessary to close the  gap, the ground state acquires a
finite magnetization, and a number of new
phenomena can appear, including critical phases, field-induced ordering,
magnetization plateaux, etc.

The system behavior at $H>H_{c}$ depends strongly on the symmetry properties
and dimensionality. Generally, if there is no axial symmetry with respect to the
field direction, the high-field phase always exhibits a tranverse staggered
long-range order (LRO) perpendicular to the applied field, independently on the
system dimensionality, and is characterized by a finite spectral gap. If the axial
symmetry is present, it cannot be spontaneously broken in a one-dimensional (1d)
system.  The high-field phase is in this case characterized by quasi-LRO (power-law
correlations), and its lowest excitations are determined by the spinon continuum.

In the 3d case, $\rm U(1)$ symmetry gets spontaneously broken and the high-field
phase is ordered but possesses a gapless (Goldstone) mode.  If one views the process
of formation of the high-field phase as accumulation of hardcore bosonic particles
(magnons) in the ground state, the ordering transition at $H=H_{c}$ can be
interpreted as the Bose-Einstein condensation (BEC) of magnons.  The idea of
field-induced BEC was discussed theoretically many times.
\cite{Affleck90-91-Sachdev+94-GiamarchiTsvelik99} The best available realisation of
such a transition was observed \cite{oosawa-jpcm,Tanaka+01,Nikuni+00} in $\rm
TlCuCl_{3}$, which can be viewed as a system of coupled $S=\frac{1}{2}$
dimers. The temperature dependence of the uniform magnetization was found
\cite{Nikuni+00} to agree qualitatively with the BEC theory predictions.

High-frequency magnetic resonance measurements
\cite{tanaka-esr,takatsu-esr} have demonstrated directly  the
field dependence of the energy gap, but were limited to the fields
$H<H_c$ and a few number of microwave frequencies.
The response in the  high-field ordered  phase of $\rm TlCuCl_{3}$ was measured in the
inelastic neutron scattering (INS)
experiments of R\"uegg et al.; \cite{Ruegg+02,Ruegg+03} the behavior of the lowest
triplet gaps as functions of field was 
successfully
described within the bond-operator mean-field theory.\cite{Matsumoto+02} In those
INS measurements only two triplet modes were observed above $H_{c}$, and the gap of
the third mode was concluded to be zero within the experimental resolution; however,
the low-energy range could not be studied in
 this experiment because of the strong field-induced magnetic
Bragg contamination below $0.25$~meV ($60$~GHz).

Very recently, high-field electron spin resonance experiments on $\rm TlCuCl_{3}$ in
a wide range of fields up to $90$~kOe were conducted, \cite{Glazkov+03} which
revealed a reopening of the gap above $H_{c}$ in the low-energy range inaccessible by
means of INS. A natural explanation would be the existence of some anisotropic
interactions explicitly breaking the $\rm U(1)$ symmetry.  The aim of the present
work is to show that the available data can indeed be described on a \emph{quantitative}
level, assuming presence of a weak exchange anisotropy in both intra-
and interdimer  interactions.

\section{Experimental summary}

 The crystals of TlCuCl$_3$ have monoclinic symmetry, with crystallographic axes $a$
and $c$ forming an angle of 96.32$^{\circ}$,  $b$ being the
twofold axis. The sample growth is described in detail in Ref.\
\onlinecite{oosawa-jpcm}. ESR spectra were taken at temperature
$1.5$~K, in the field range from $0$ to $90$~kOe, using a set of
home-made microwave spectrometers with transmission type cavities
and a superconducting magnet. Single crystals with the volume of
$20\div50$mm$^{3}$ were used. During the experiments crystals were
mounted in the following orientations with respect to the magnetic
field: $H||[010]$ (i.e., parallel to the $b$ axis),
$H\perp(10\overline{2})$ and $H||[201]$. The $[201]$ direction
forms an angle of $51.7^{\circ}$ with the $a$ axis, see Fig.\
\ref{fig:axes}. The detailed description of the ESR spectra is
reported in Ref.\ \onlinecite{Glazkov+03} and here we will just
summarize the results.

 The ESR signal corresponding to the transitions from the ground state
to the lowest $q=0$ excited state is clearly visible both below and
above the critical field. For $H<H_{c}$, such transitions would be
forbidden for an ideal isotropic system.
 The main feature of this signal
is reopening of the gap above the critical field, which can be
directly observed for all three mutually perpendicular field
directions. Besides the ground state transitions, transitions between
Zeeman-split components of the thermally activated triplet were
observed below the critical field. The analysis of the field
dependence of thermally activated transitions at $H<H_{c}$
has also suggested presence of a finite zero-field splitting of the
triplet.\cite{Glazkov+03}

\section{The effective model}

The dynamics of a 3d coupled anisotropic $S=\frac{1}{2}$ dimer system
in a wide range of fields can be described within the effective field
theory \cite{K96,Zheludev+03b} which may be viewed as a continuum
version of the bond boson approach.\cite{SachdevBhatt90,Matsumoto+02}
It is based on introducing dimer coherent states \cite{K96}
\begin{equation}
\label{dimer-wf}
|\vec{A},\vec{B}\rangle=(1-A^2-B^2)^{1/2} 
|s\rangle+\sum_j (A_{j}+iB_{j})|t_j\rangle,
\end{equation}
where $|s\rangle$  and $|t_j\rangle$, $j=(x,y,z)$ are the singlet and three
triplet states,\cite{SachdevBhatt90} and $\vec{A}$, $\vec{B}$
are real vectors related to the magnetization
$\vec{M}=\langle\vec{S}_1+\vec{S}_2\rangle$ and sublattice magnetization
$\vec{L}=\langle\vec{S}_1-\vec{S}_2\rangle$
of the spin dimer:
\begin{eqnarray}
&&\vec{M}=2(\vec{A}\times\vec{B}),\quad
\vec{L}=2\vec{A}\sqrt{1-A^2-B^2}.
\label{ML}
\end{eqnarray}

\begin{figure}[tb]
\includegraphics[width=50mm]{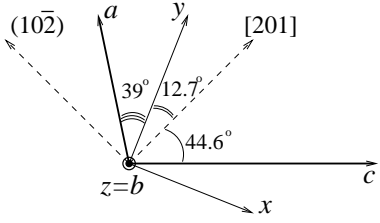}
\caption{\label{fig:axes}  Schematic picture of the anisotropy
axes in $\rm TlCuCl_{3}$.}
\end{figure}

In $\rm TlCuCl_{3}$, there are two types of chains of spin dimers running along the
crystallographic $a$ axis. 
In the high-field phase the staggered order
$\vec{L}$ alternates between the two different types of chains. \cite{Tanaka+01}
The magnetization
$\vec{M}$ should be uniform, 
which leads us to the following
ansatz:  
\[
\vec{A}_{\lambda\mu\nu}= (-1)^{\mu+\nu}\vec{\varphi}(\vec{R}_{\lambda\mu\nu})
\quad
\vec{B}_{\lambda\mu\nu}= (-1)^{\mu+\nu}\vec{\eta}(\vec{R}_{\lambda\mu\nu}),
\]
where $\vec{R}_{\lambda\mu\nu}=
\lambda\vec{e}_{1}+\mu\vec{e}_{2}+\nu\vec{e}_{3}$ is the radius-vector
labeling the dimers, and the vectors $\vec{e}_{1,2,3}$ are in the
following way connected to the lattice vectors $\vec{a}$, $\vec{b}$, $\vec{c}$:
\[
\vec{e}_{1}=\vec{a},\quad \vec{e}_{2}=(\vec{b}-\vec{c})/2,\quad
\vec{e}_{3}=(\vec{b}+\vec{c})/2.
\]

The topology of exchange paths in $\rm TlCuCl_{3}$ is known
rather well. \cite{excitations,Matsumoto+02} We assume additionally a weak
orthorhombic anisotropy (the simplest type compatible with
the crystal symmetry) in intra- as well as in interdimer
exchange,\cite{note1} so that
instead of one intradimer exchange constant $J$ one has 
a vector $\vec{J}=\{J_{x},J_{y},J_{z} \}$, etc. Each dimer
has an inversion center, which excludes the intradimer
Dzyaloshinskii-Moriya (DM) interaction  (but it remains possible
between the dimers); for simplicity, we 
neglect  the DM
interaction in the present treatment.

Passing to the continuum, one obtains the  Lagrangian
\begin{eqnarray}
\label{Leff}
{\mathcal L}&=& -2\hbar \vec{\eta}\cdot\partial_{t}\vec{\varphi} -\frac{1}{2}
\sum_{i=x,y,z} \beta_{i} \sum_{j=1,2,3}
 [(\vec{e}_{j}\cdot\vec{\nabla})\varphi_{i}]^{2} \\
&-&\sum_{i}\{
m_{i}\varphi_{i}^{2}
+\widetilde{m}_{i}\eta_{i}^{2} \}
+2\vec{h}\cdot(\vec{\varphi}\times\vec{\eta})
-V(\vec{\varphi},\vec{\eta}),\nonumber
\end{eqnarray}
where $m_{i}=\widetilde m_{i}-\beta_{i}$ and $\widetilde
m_{i}=\frac{1}{4}|\epsilon_{ijn}|
(J_{j}+J_{n})$; we denote
$h_{i}=\sum_{j}g_{ij}\mu_{B}H_{j}$, where $g$ is the gyromagnetic tensor and
$\vec{H}$ is the  magnetic field. 
Effective interdimer couplings $\beta_{i}$ are in the following
way connected with the microscopic couplings (the notation of
Ref.\ \onlinecite{excitations} is used):
\begin{equation} 
\label{beta}
\vec{\beta}=\vec{J}'_{(100)} +\vec{J}'_{(201)} +2\vec{J}_{(1{1\over2}{1\over2})}
 -2\vec{J}_{(100)}-2\vec{J}'_{(1{1\over2}{1\over2})}.
\end{equation}

We will assume that we are not too far above the
critical field, so that the magnitude of the triplet component is small, i.e.,
$\varphi,\eta\ll1$. 
Then, retaining only the fourth-order terms in
the interaction $V$ in (\ref{Leff}), one obtains
\begin{eqnarray} 
\label{V4} 
&&V=\sum_{i}\beta_{i}\varphi_{i}^{2}\vec{\varphi}^{2} 
+\sum_{ij}(\gamma_{ij}\eta_{i}^{2}\varphi_{j}^{2}
+\lambda_{ij} \varphi_{i}\varphi_{j}\eta_{i}\eta_{j}),\nonumber\\
&&\gamma_{ij}=\beta_{i}-\lambda_{ij},\quad
\lambda_{ij}=-\sum_{l}|\epsilon_{ijl}|\alpha_{l},
\end{eqnarray}
where
$\alpha_{i}$ is given by a different combination of couplings,
\begin{equation} 
\label{alpha} 
\vec{\alpha}=\vec{J}'_{(100)} +\vec{J}'_{(201)} +2\vec{J}_{(1{1\over2}{1\over2})}
 +2\vec{J}_{(100)}+2\vec{J}'_{(1{1\over2}{1\over2})}.
\end{equation}

The further derivation is similar to that given in Ref.\
\onlinecite{Zheludev+03b}.  Integrating out the field $\vec{\eta}$ 
 results in the relation
\begin{eqnarray}
\label{B}
&&\eta_{i}=Q_{ij}F_{j},\quad \vec{F}= -\hbar\partial_{t}\vec{\varphi}
+(\vec{h}\times\vec{\varphi}) \nonumber\\
&& Q_{ij}={\delta_{ij}\over \widetilde{m}_{i} }
-\sum_{k}\gamma_{ki} {\delta_{ij}\varphi_{k}^{2}\over \widetilde{m}_{i}^{2}}
-\lambda_{ij}{\varphi_{i}\varphi_{j}\over \widetilde{m}_{i}\widetilde{m}_{j}} .
\end{eqnarray}
and yields the
effective Lagrangian depending on $\vec{\varphi}$ only:
\begin{eqnarray}
\label{Leff-a}
{\mathcal L}&=& \sum_{i}{1\over \widetilde{m}_{i}}
\hbar^{2}(\partial_{t} \varphi_{i})^{2} 
-\frac{1}{2}
\sum_{i=x,y,z}\beta_{i} \sum_{j=1,2,3}
 [(\vec{e}_{j}\cdot\vec{\nabla})\varphi_{i}]^{2} \nonumber\\
&-&\sum_{i}2{\hbar\over \widetilde{m}_{i}} (\vec{h}\times\vec{\varphi})_{i}
\partial_{t} \varphi_{i} -U_{2}(\vec{\varphi})
  -U_{4}(\vec{\varphi},\partial_{t}\vec{\varphi}),
\end{eqnarray}
with the quadratic and quartic interaction given by
\begin{eqnarray}
\label{U2-4}
 U_{2}&=&
m_{i}\varphi_{i}^{2}-{1\over\widetilde{m}_{i}}(\vec{h}\times\vec{\varphi})_{i}^{2},\\
 U_{4}&=& 
\sum_{i} \beta_{i}\varphi_{i}^{2}\vec{\varphi}^{2}
+\sum_{ij}\gamma_{ji}{\varphi_{j}^{2}F_{i}^{2}\over\widetilde{m}_{i}^{2}}
+\sum_{ij}\lambda_{ij}
{\varphi_{i}\varphi_{j}\over \widetilde{m}_{i}\widetilde{m}_{j}} F_{i}F_{j}.\nonumber
\end{eqnarray}
We  analyze the obtained field theory at the mean-field
(zero-loop) level;
this is formally justified for weak
interdimer coupling, while in $\rm TlCuCl_{3}$ 
$\beta_{i}$ and $\widetilde{m}_{i}$ are of the same order. 
However, 
a similar approximation is used in the bond-boson approach which
successfully describes the INS data on the magnon
dispersion in this material.\cite{Matsumoto+02} As we will see
later, this simplified treatment yields reasonable results in the present
case as well.

The staggered order parameter $\vec{\varphi}^{(0)}$ (static value of
$\vec{\varphi}$) is zero below
$H_{c}$, and above the critical field it is determined as the nontrivial
solution of the equations
\begin{equation}
\label{eq-A0}
\sum_{j}\Omega_{\beta j}\varphi_{j} +\sum_{imn}\Lambda_{\beta i,mn} \varphi_{i}
\varphi_{m}\varphi_{n}=0, 
\end{equation}
where the matrices $\mathsf{\Omega}$, $\mathsf{\Lambda}$ are defined as
\begin{eqnarray}
\label{mat1}
&& \Omega_{ij}=m_{i}\delta_{ij}
-\sum_{kln}\epsilon_{ikn}\epsilon_{jln}\frac{h_{k}h_{l}}{\widetilde{m}_{n}}, 
\nonumber\\
&& 
\Lambda_{ij,mn}=\Gamma_{ij,mn}+\Gamma_{mn,ij},\\
&& \Gamma_{ij,mn}=\beta_{i} \delta_{ij}\delta_{mn}
+\delta_{ij}
\sum_{kls}{\gamma_{ik}\over\widetilde{m}_{k}^{2}}\epsilon_{klm}\epsilon_{ksn}
h_{l}h_{s}\nonumber\\ 
&& \qquad +\lambda_{ij}\sum_{kl}
\epsilon_{ikm}\epsilon_{jln}\frac{ h_{k}h_{l} }{
\widetilde{m}_{i}\widetilde{m}_{j} }.\nonumber
\end{eqnarray}

Linearizing the theory around 
$\vec{\varphi}=\vec{\varphi}^{(0)}$, one finds  
the magnon energies $E$ depending on the field $\vec{H}$ and  the
wave vector
$\vec{q}$ as  real roots
of the secular equation
\begin{equation}
\label{secular}
\det( \mathsf{M} -E^{2}\mathsf{G}
-iE\mathsf{C})=0,
\end{equation}
where $ \mathsf{G}_{ij}=Q_{ij}(\vec{\varphi}^{(0)})$ 
and the matrix $\mathsf{M}$ is given by
\begin{eqnarray}
\label{mat2}
\mathsf{M}_{ij}&=&\Omega_{ij} +{1\over2}\beta_{i}\delta_{ij}  \sum_{k}
(\vec{q}\cdot\vec{e}_{k})^{2} \\
&+& \sum_{mn} \varphi_{m}^{(0)} \varphi_{n}^{(0)}
(\Lambda_{ij,mn}+\Lambda_{im,jn}+\Lambda_{in,mj}).\nonumber
\end{eqnarray}
The antisymmetric matrix $\mathsf{C}$ can be written as
\begin{eqnarray} 
\label{mat3} 
\mathsf{C}_{ij}&=&\Big(\frac{1}{\widetilde{m}_{i}} +\frac{1}{\widetilde{m}_{j}}
\Big) \sum_{k}\epsilon_{ijk} h_{k} 
-\sum_{kl}\varphi^{(0)}_{k} \varphi^{(0)}_{l} \\
&\times& (S_{ikl,j}+S_{kil,j} +S_{kli,j}
-S_{jkl,i} -S_{kjl,i}-S_{klj,i}),\nonumber
\end{eqnarray}
where $S_{ikl,j}$ is defined as follows:
\begin{equation} 
\label{S} 
S_{ikl,j}=\gamma_{kj}\delta_{kl}\frac{1}{\widetilde{m}_{j}^{2}}
\sum_{r}\epsilon_{ijr}h_{r} 
+\frac{\lambda_{kl}\delta_{kj}}{\widetilde{m}_{l}\widetilde{m}_{j}}
\sum_{r}\epsilon_{ilr}h_{r}.
\end{equation}

To proceed, one has to fix the  principal anisotropy axes. As
suggested in Ref.\ \onlinecite{Glazkov+03}, for symmetry reasons one of them should
coincide with the crystallographic $b$ axis; this axis we will denote $z$.  Another
one, which we label $y$, should be the axis along which the spin ordering
occurs at $\vec{H}\parallel \vec{b}$; according to Ref.\ \onlinecite{Tanaka+01}, it
lies in the $(ac)$ plane and forms the angle of $39^{\circ}$ with the $a$ axis, see
Fig.\ \ref{fig:axes}.

\begin{figure}[tb]
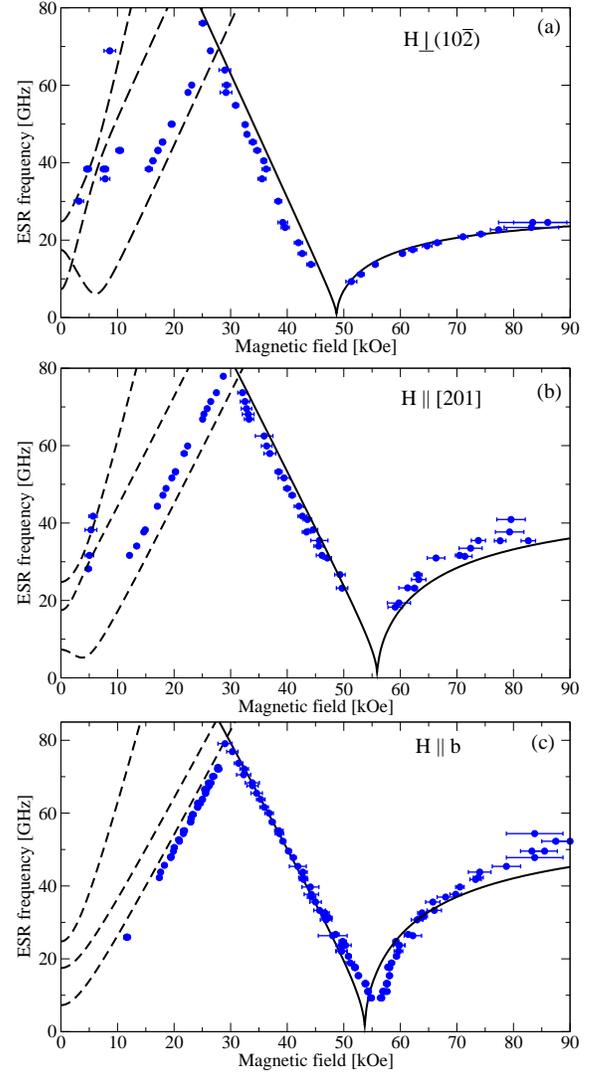

\includegraphics[width=75mm]{newc102.eps}

\includegraphics[width=75mm]{newc201.eps}

\includegraphics[width=75mm]{newcB.eps}
\caption{\label{fig:fit} ESR frequency-field
dependencies\cite{Glazkov+03} taken at $T=1.5$~K for different field orientations, in
comparison with the  present theory.  Solid and
dashed lines correspond to the ground state and thermally activated
transitions, respectively.}
\end{figure}

The best fit to the entire set of the experimental 
frequency-field dependencies is obtained with 
\begin{equation} 
\label{fit} 
\vec{\beta}=\{ 5.628, 5.638, 5.635 \} , 
\quad
\vec{\widetilde{m}}=\{ 5.725, 5.711, 5.714 \} .
\end{equation}
Here all values are given in meV, the absolute error for the \emph{anisotropic} part of $\vec{\beta}$,
$\vec{\widetilde{m}}$ is about $10^{-3}$, but for
the \emph{isotropic} part it is larger, about $0.1$: 
the results are not very sensitive to a shift
of $\beta_{i}$ and $\widetilde{m}_{i}$ by the same value.

We assumed the $g$-factor to be diagonal in the chosen anisotropy axes and took
$g_{xx}=2.29$, $g_{yy}=g_{zz}=2.06$.  
The theoretical curves are shown in Fig.\
\ref{fig:fit} in comparison to the experimental results.
All gaps occur at $\vec{q}=0$.
 Solid lines show the lowest magnon
gap, and the dashed ones correspond to the transitions
between three branches of the magnon triplet at $q=0$.
The theory is in a good agreement with the low-temperature ESR
results.

We were not able to estimate $\vec{\alpha}$ from our fit, since 
the results are insensitive to the exact value
of $\alpha_{i}$: indeed,
$\vec{\alpha}$ enters only the
$\eta^{2}\varphi^{2}$-type part of the
interaction (\ref{V4}), and in the ordered phase
$\eta\propto (h/J)\varphi$, which suppresses
the contribution of such terms   in the relevant field
range $h/J< 0.1$.
Available estimates\cite{excitations}
of ${J}_{(100)}$, ${J'}_{(1{1\over2}{1\over2})}$ suggest that
 $\alpha_{i}-\beta_{i}\approx -0.92$~meV, which has been
assumed for the curves presented in Fig.\
\ref{fig:fit}.

Above the critical field the lowest gap opens as $E_{g}\simeq C \sqrt{h^2-h_c^2}$,
e.g., 
for $\vec{H}\parallel z$ and weak anisotropy, 
\[
 C^{2}\simeq  2 (\widetilde{m}_{y}/m_y)
 (\widetilde{m}_{y} m_x - \widetilde{m}_{x} m_y)
/(\widetilde{m}_{x} +\widetilde{m}_{y})^2\simeq 0.16,
\]
where the weak contribution of $\gamma_{ij}$, $\lambda_{ij}$ is neglected.

\section{Discussion}

Our analysis suggests that the anisotropy in
intradimer interactions $\widetilde{m}_{i}$ as well as in inter-dimer
couplings $\beta_{i}$ is very small and does not exceed one percent,
which is plausible for the exchange anisotropy.\cite{note} From Eq.\ 
(\ref{fit}) one may get the impression that the intradimer exchange
has a different sign of anisotropy.  However, $\vec{\widetilde{m}}$
given in (\ref{fit}) corresponds to the physical intradimer coupling
$\vec{J}=\{5.700,5.728,5.722\}$, so that the anisotropy has the same
character for inter- and intradimer exchange:
$y$ is the easy axis, and $z=b$
is the intermediate  axis.

\begin{figure}[t]
\includegraphics[width=78mm]{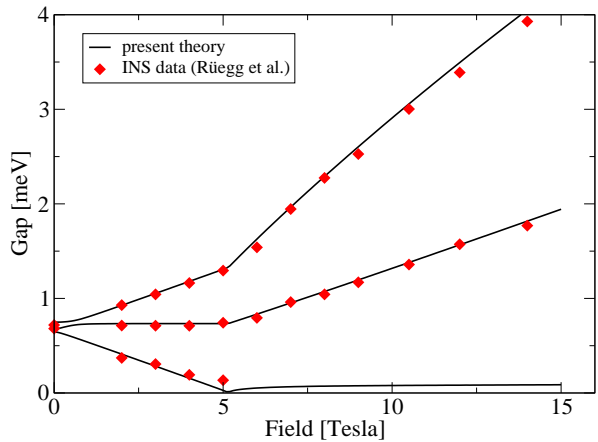}
\caption{\label{fig:ins} INS results of R\"uegg et al.\cite{Ruegg+02,Ruegg+03}
taken at $T=1.5$~K, compared with the present theory.  In this geometry 
 ($\vec{H}$ is perpendicular to the plane
defined by $(010)$ and $(104)$ vectors) the predicted
lowest gap
 stays below $0.09$~meV at $H>H_{c}$, i.e., beyond the available
experimental resolution.  }
\end{figure}

According to our calculations, the staggered
order parameter $\vec{\varphi}^{(0)}$ is directed along the $y$ axis for
$\vec{H}\parallel \vec{b}$ and $\vec{H}\perp (10\overline{2})$ (with a
tiny deviation from the $(yz)$ plane in the latter case);
for $\vec{H}\parallel [201]$, we
predict that $\vec{\varphi}^{(0)}\parallel\vec{b}$. The
magnitude of the order parameter at $H=90$~kOe is about $0.1$ of the
saturation value, for all three field geometries.  Those predictions
agree with the existing  results of elastic neutron experiments
\cite{Tanaka+01} for $H\parallel b$, and can be tested for other
orientations. 
Our results are  consistent with the
INS data\cite{Ruegg+02,Ruegg+03} as well, see Fig.\ \ref{fig:ins}.

It is worthwhile to note that the same conclusion on the character of
the anisotropy ($y$ and $b$ being the easy and the second easy axis)
was reached in recent ESR studies\cite{Shindo+unpub} of the
impurity-induced order in  $\rm Mg$-doped $\rm TlCuCl_{3}$.

We are grateful to A. Furusaki, H.-J. Mikeska, Ch.~R\"uegg,
A. I. Smirnov, and K. Totsuka for fruitful discussions.  This
work is supported in part by Grant I/75895 from the
Volkswagen-Stiftung, by Grant No.\ 03-02-16579 from Russian Foundation
for Basic Research, by INTAS Grant No.\ 04-5890, and by Grant-in-Aid
for Scientific Research on Priority Areas ``Field-Induced New Quantum
Phenomena in Magnetic Systems''.


\begin{thebibliography}{10.}
\bibitem{note1} We have checked that introducing anisotropy into only one
(intra- or interdimer) set of couplings is not sufficient for the
description of the experimental data of Ref.\ \protect\onlinecite{Glazkov+03}.
\bibitem{note} There is another solution, $\vec{\widetilde{m}}=\{ 5.500,
  5.793, 5.500 \}$ and $\vec{\beta}=\{5.401, 5.716, 5.420 \}$, which
yields an even better fit but  corresponds to unrealistically large
 anisotropy (about 10\% in the intradimer echange)  and was hence discarded.
\end{thebibliography}
\end{document}